\documentstyle[aps,prl,epsf,floats]{revtex}
\begin{document}
\ifpreprintsty \else
\twocolumn[\hsize\textwidth\columnwidth\hsize\csname@twocolumnfalse%
\endcsname \fi

\draft
\title{Friction, order, and transverse pinning 
of a two-dimensional elastic lattice \\under periodic and impurity potentials}
\author{Takaaki Kawaguchi}
\address{Department of Technology, Faculty of Education, Shimane 
University\\
1060 Nishikawatsu, Matsue 690-8504, Japan}
\author{Hiroshi Matsukawa}
\address{Department of Physics, Osaka University\\
1-16 Machikaneyama Toyonaka, Osaka 560-0043, Japan}
\maketitle
\begin{abstract}
Frictional phenomena of two-dimensional elastic lattices are studied
numerically based on a two-dimensional Frenkel-Kontorova model
with impurities.
It is shown that impurities can assist the depinning.
We also investigate anisotropic ordering and transverse pinning 
effects of sliding lattices,
which are characteristic of the moving Bragg glass state and/or 
transverse glass state.
Peculiar velocity dependence of the transverse pinning is observed 
in the presence of both periodic and random potentials and discussed 
in the relation with growing order and discommensurate structures.

\end{abstract}
\pacs{ }
\ifpreprintsty \else
] \fi              

\narrowtext
%
Static and dynamic properties of an elastic or plastic media, subject to an 
external force and under  periodic or random potentials, have attracted much 
attention recently in relation to the vortex lattice in type-II 
superconductors, charge and spin density waves, Wigner crystals, the 
interface growth, the dry friction between two solids 
and so on\cite{tribo,fric,gia2,dou,bal,moon,ryu}.
They are pinned by those potentials under a small external force and start 
a kind of sliding motion above a certain threshold force.
Various peculiar phenomena are predicted and/or observed in these systems.
Most theoretical or numerical works study, however, systems under only 
random potentials or those under only periodic potentials.
The latter ones are investigated as models of vortex lattice with intrinsic 
pinning, commensurate density waves, the friction between atomically flat 
surfaces, etc.
In actual experiments of those systems, a finite amount of 
randomness will exist.
Theoretical or numerical works of systems under random potentials in addition to 
periodic potentials are indispensable.
The behavior of friction in the presence of impurities in addition to the 
periodic potential has been studied, based on a simplified one-dimensional model, 
numerically\cite{Vinokur,kawa1} and theoretically\cite{Yumoto}.
The interfaces where the friction acts are, however, in most cases two-dimensional, and 
the frictional properties are expected to strongly depend on the dimensionality of 
the system.
In this paper, we investigate numerically the behavior of friction based 
on a two-dimensional elastic lattice in the presence of impurity 
and incommensurate periodic potentials \cite{ECRY}.
The model is also applicable to the vortex 
lattice in thin films, and so on.
We show that the maximum static frictional force in the presence of 
impurities can be lower than that in the absence of them.
This is caused by the frustration between impurities and the periodic 
potential.
We also investigate the transverse pinning\cite{gia2,dou,bal}.
When we apply an external force along the transverse direction to the sliding lattice, 
the sliding velocity has a finite transverse component only for the transverse 
external force that is greater than a finite critical value.
This transverse pinning is predicted in the case under random potentials 
alone and the manifestation of a quasi long-range order (QLRO) 
along the transverse direction.
We calculated the structure factor of the sliding lattice, which show 
clearly such transverse QLRO.
The velocity dependence of the transverse threshold force can show 
nonmonotonic behavior in the presence of both periodic and random potentials.
The velocity dependence of the kinetic frictional force is also presented.
They are discussed 
in the relation with the QLRO and discommensurate structures of sliding lattice.

As a simple model which describes an interface between two solids,
we consider here a two-dimensional Frenkel-Kontorova (FK) model \cite{fk} 
with impurities.
The model consists of a square lattice of atoms interacting with each other 
via a harmonic force on a substrate.
The periodic potential resulting  from the substrate is represented by a 
two-dimensional sinusoidal form,
$ U_{P}({\bf r})= K \left[ \sin\left( \frac{2\pi}{C_p}x \right)+ 
\sin\left( \frac{2\pi}{C_p}y \right) \right] $.
The impurity potential $U_{I}({\bf r})$ is expressed as,
$ U_{I}({\bf r})=-\frac{W}{2}\sum_{i}^{N_I} \exp{\left(-4({\bf r}-{\bf 
R}_i)^2 \right)} $, where
$N_I$ is the number of impurities and ${\bf R}_i$ the coordinate of 
the $i$th impurity
distributed randomly on the substrate.
The overdamped equation of motion of each atom is given by
\begin{eqnarray}
\dot{u}^\alpha_{i,j}
&=&u^\alpha_{i+1,j}+u^\alpha_{i-1,j}+u^\alpha_{i,j+1}+u^\alpha_{i,j-1}-4u^\alpha_{i,j}  
\nonumber \\
&& +F^\alpha_{P}({\bf u}_{i,j})+F^\alpha_{I}({\bf u}_{i,j}) + 
F^\alpha_{ex},
\end{eqnarray}
where $u_{i,j}$ is the position of the atom with index numbers $i$ and $j$,
and the superscript $\alpha$ stands for x and y.
In the above, $F_{P}({\bf r}) =- \nabla U_{P}({\bf r})$ and $F_{I}({\bf r}) 
=- \nabla U_{I}({\bf r})$ are the force from the periodic 
potential and that from impurities, respectively, and $F_{ex}$ 
is the external force.
We apply $F_{ex}$ along the x-axis, except in the study of the transverse pinning.

In this paper we consider an incommensurate system
and choose the ratio between the mean atomic spacing of the lattice, $c_a$,
and the period of the periodic potential, $c_p$,
as the golden mean ratio:
$c_a /c_p = \sqrt{5} +1 /2=1.618\cdots$ (we set $c_p=3$ and 
$c_a=1.618\cdots\times c_p$).
The number of atoms in the samples $N$ is $55^2$, $89^2$, or $144^2$, and
the number of impurities $N_I$ is set so that the impurity density $N_I/(N c_a^2)$ 
is fixed at about 0.21.
Numerical simulations are performed using a Runge-Kutta formula
under periodic boundary conditions.
Hereafter, $v^\alpha$ is the velocity averaged over space and time.

Note that the periodic boundary conditions force the system as a whole to 
remain incommensurate.
The results obtained here are, however, expected to be semiquantitatively the 
same for a large system with open boundary conditions which is originally 
incommensurate, such as experimental systems.
A kind of discommensurate structure with commensurate domains separated by 
defects plays an essential role for the present results.
Such a structure appears also for the large system with open boundary 
conditions.
Hence we can simulate the large system with open boundary conditions by a 
small system with periodic boundary conditions.
%

First, we investigate pinning properties of static states ($v^x = v^y=0$) by evaluating
the depinning threshold force  $F_c$ for various values of $K$ and $W$.
According to the scaling theory of the pinning of density-waves \cite{fuku},
$F_c \propto W^{\frac{4}{4-d}}$ for weak pinning and $F_c \propto W$ for 
strong pinning,
where $d$ is the spatial dimension \cite{matsu1}.
On the other hand, in the presence of only the periodic potential, 
the lattice is pinned due to the occurrence of the Aubry transition \cite{aubry} 
for $K>K_c$, where $K_c\approx 0.23$ in the present system.
We can expect that these two pinning mechanisms compete with each other,
as observed in one-dimension \cite{Vinokur,kawa1,Yumoto}.

In Fig. 1 the depinning threshold forces $F_c$'s are plotted against $W$.
In the absence of the periodic potential, $K=0$,
the weak pinning behavior, $F_c \propto W^2$, is clearly observed in a 
small $W$ regime, $W\ll 10$,
and then, with increasing $W$, the $F_c-W$ curve gradually approaches the 
strong pinning behavior,
$F_c \propto W$.
When the periodic potential with a finite $K(>K_c)$ is present in 
addition to impurities, however, 
the weak pinning behavior is modified especially in the small $W$ regime.
In this case, the pinning is governed dominantly by the periodic potential
and the effect of impurities becomes only a weak perturbation to the 
pinned state.
As the strength of impurity potential $W$ is increased,
the pinned state crosses over to the weak or strong pinning regime.
In the crossover regime, an interesting behavior is observed, that is, 
$F_c$ takes a minimum value.
In this regime, impurities diminish the pinning effect of the periodic 
potential.
In other words, impurities assist the depinning.
Such a behavior is observed in a one-dimensional system \cite{Vinokur} 
and caused by the frustration between impurities and the periodic 
potential\cite{Yumoto}. 
%

Next we focus on sliding states.
We consider here cases in which the impurity potential is much stronger than
the periodic potential, $W\gg K$, for which
the static states  are pinned dominantly by the impurity 
potential.
To clarify the lattice order,
we calculate the structure factor in steady sliding states:
$
S({\bf k})=\frac{1}{N^2}\sum_{i,j}\exp{i{\bf k}\cdot({\bf u}_i-{\bf u}_j)}.
$
Typical results of $S({\bf k})$ at some sliding velocities are shown in 
Fig. 2.
In a low velocity regime (Fig. 2(a)) Bragg peaks at $k_x\neq0$ disappear.
This state corresponds to the moving transverse glass (MTG) state 
discussed
in Refs.\cite{gia2,dou,bal}, where QLRO only along the transverse direction exists.
So far, the MTG states have been observed in numerical simulations of 
vortex lattices, where
the longitudinal positional long-range order is destroyed by
the dislocations with Burger's vectors parallel to the moving direction\cite{moon}.
Such dislocations are produced dynamically by the random force
resulting from the interaction with impurities.
Also, for the present harmonic lattice case
such disordered sliding states appear.
Even in small $W$ cases, when the velocity is sufficiently small,
the MTG or a MTG-like anisotropic state can be observed.
The appearance of the MTG state would be a common feature
for two-dimensional driven lattices under random potentials.
As the driving force is increased, Bragg peaks at $k_x\ne 0$ appear 
and grow gradually,
i.e., the MTG state crosses over to the moving Bragg glass(MBG) state, 
where QLRO exists along both directions.
In Fig. 2(b), 
the satellite Bragg peaks on the $k_x$ axis ($k_x\ne 0$ and $k_y=0$) are 
slightly higher than
those on the $k_y$ axis ($k_x=0$ and $k_y\ne 0$).
The width of the peaks is also anisotropic and broadens in the $k_y$ 
direction.
The anisotropy of the MBG survives in the high velocity regime.
Note here that, in two-dimensional systems,
the change from MTG to MBG would occur as a crossover \cite{dou,bal}.

Figure 3 shows the kinetic frictional force, $F_{kin}$, as a function of 
the velocity $v_x$
for $W=10$ and $K=0$ and $1.1$.
For comparison, $F_{kin}$ for $W=0$ and $K=1.1$ is plotted in the inset.
In the present model $F_{kin}$ is given by
$F_{kin}=F^x_{ex}-v_x =
- \frac{1}{N}\sum_{i,j}\left\langle F_P( u_{i,j})+F_I( 
u_{i,j})\right\rangle_{t}
$,\cite{matsu2} 
where the contribution from the viscous term in Eq. (1) is subtracted 
and the angle bracket means temporal average\cite{matsu2}.
In the low velocity regime, $v_x \ll 10$, the periodic potential plays no role and 
the velocity dependence of $F_{kin}$ is weak.
In the zero velocity limit, both $F_{kin}$'s agree with $F_{c}$'s for 
$K\rightarrow 0$ in Fig. 1.
The weak velocity dependence of $F_{kin}$ may relate to the fact that the impurity force 
acting on two-dimensional CDW's is independent from the sliding velocity.\cite{matsu1} 
As $v_x$ increases, the velocity dependence crosses over to the behavior, 
$F_{kin} \propto v_x^{-1}$.
When $v_x$ is increased further, the dependence on the periodic potential amplitude, 
$K$, appears.
This behavior is consistent with the perturbation theory \cite{matsu1,soko,prsn,kawa2} and 
the growing order along the longitudinal direction observed in Fig. 2.
The main contribution to the kinetic frictional force comes from 
the dynamical lattice deformation 
along the longitudinal direction, $u^x$.\cite{gia2,dou}
As discussed later, the periodic potential becomes effective to the kinetic frictional force 
only when the characteristic length of $u^x$ grows sufficiently.  
In the low velocity regime, $u^x$ fluctuates so strongly due to the random potentials. 
Then, the characteristic length is small and 
the effects of the periodic potential almost vanishes.
When $v_x$ increases, QLRO of $u^x$ grows. 
Hence the fluctuation is suppressed, and then the lowest-order perturbation theory works.
It is easily shown by the perturbation theory that 
the kinetic frictional force is given by the sum of the contribution 
from the periodic potential and that from the impurity potential.
Both contributions have the same velocity dependence in the large velocity regime, 
they are proportional to $v_x^{-1}$.\cite{matsu1,soko,prsn,kawa2}
As a result, in the large velocity regime, the strength of the kinetic frictional force 
depends on the amplitude of the periodic potential, 
but its velocity dependence in that regime does not depend on it.
The periodic-potential dependence of the kinetic frictional force is 
the manifestation of the crossover from MTG to MBG.
%

We next investigate the transverse response of sliding lattices.
Giamarchi and Le Doussal developed a kind of the scaling theory on the MBG state
for $d$-dimensional lattices under impurity potentials.\cite{gia2,dou}
They clarified that if QLRO exists along the transverse direction,
then a certain pinning force works in that direction even in sliding states.
According to their analysis, the transverse depinning threshold force 
$F_c^y$ is given by
\begin{eqnarray}
F_c^y  \sim \left( \frac{W^2 n_i}{v_x}\right)^2 ,
\end{eqnarray}
in the large velocity regime, where $n_i$ is the impurity density.
For numerical evaluation of $F_c^y$,
transverse force $F_y$ is applied to a sliding lattice driven by 
longitudinal force $F_x$,
and then transverse velocity of the lattice is monitored in steady states.
Figure 4 shows $F_c^y$ against the longitudinal velocity $v_x$ for 
$W=10$ and several values of $K$ much smaller than $W$.
Finite transverse threshold force is clearly observed.
For $K=0$, $F_c^y$ shows a crossover behavior.
In the  low $v_x$ regime, $v_x \leq 3$, $F_c^y$ is almost constant.
When $v_x$ is increased, $F_c^y$ becomes proportional to  $v_x^{-2}$.
Such a crossover  behavior is also predicted by Giamarchi and Le Doussal around  
the velocity $v_x \simeq F_c$ in the present unit, 
which is consistent with the present result.
In the limit of vanishing velocity, $v_x\rightarrow 0$, $F_c^y$ is 
much smaller than the depinning threshold force in Fig. 1, $F_c$. 
Such reduction of the transverse pinning force in comparison with
the longitudinal one may come from kinematical effects.
For finite $K$'s, i.e., in the presence of the periodic potential,
peculiar velocity dependence appears.
For finite but small values of $K$, $K=0.6$ and $0.7$ in Fig. 4, 
$F_c^y$ shows a similar behavior to that for 
$K=0$ in the low and intermediate velocity regimes, and
seems to follow the scaling relation.
As $v_x$ increases further, however, $F_c^y$ takes a minimum value at certain velocities 
and then increases.
The velocity at which $F_c^y$ is minimum decreases with increasing $K$.
For larger values of $K$, $F_c^y$ is constant in the low velocity regime and 
then increases monotonically.
In the large velocity limit, 
$F_c^y$ approaches the depinning threshold field $F_c$ in Fig. 1 
in the absence of impurity, $W=0$, for any values of $K$.
This means that, in the transverse direction,
sliding lattices in the large velocity limit feel 
only the transverse barrier by the periodic potential $U_p$.

These behaviors are explained as follows.
The transverse pinning is essentially related to 
the transverse lattice deformation $u^y$.\cite{gia2,dou}
In the case of weak periodic potential and the low velocity regime, the pinning effect 
mainly comes from impurities.
The characteristic length of $u^y$ along the longitudinal direction is small.
It is to be noted that the discommensurate structure is indispensable for the pinning 
by the periodic potential in  incommensurate systems, which are investigated here.
The discommensurate structure of $u^y$  is responsible for the transverse pinning 
by the periodic potential.
There are many discommensurate structures with different length scales 
for a given boundary condition.
They are regarded as excited states from the ground-state discommensurate structure and 
the state with larger length scale has less excitations and larger pinning energy.
When one of the characteristic lengths of $u^y$ 
along the longitudinal and transverse directions 
are much smaller than a typical length scale of the discommensurate structure,  
which will be the order of the neighboring discommensuration distance, 
the resultant discommensurate structure has much excitation and a small pinning energy, and 
then the effect of the periodic potential to the transverse pinning almost vanishes.
Hence  that effect vanishes in the low velocity regime.
As the sliding velocity increases, the characteristic lengths of $u^y$ along both directions grow.
The discommensurate structures with larger length scales can be realized there.
They have a larger pinning energy, and then the effect on the transverse depinning threshold field 
is strengthened.
In the large velocity limit, the state has the ground-state discommensurate structure 
in the absence of impurities and the transverse depinning threshold field approaches 
the depinning threshold field $F_c$ in that case.
The velocity scale where the periodic potential becomes effective to the transverse pinning 
should depend on its amplitude, and is large for the weak potential.
As a result, in the case of a small value of $K$, $F_c^y$ decreases with increasing $v_x$ and 
then increases as the periodic potential becomes effective.
When $K$ is increased, that velocity scale decreases, and then the velocity 
at which $F_c^y$ is minimum also decreases.
Further increase of $K$ makes the velocity scale smaller than the crossover velocity of 
the impurity transverse pinning, and then $F_c^y$ increases monotonically 
with the sliding velocity.
This scenario is consistent with all the results presented in this paper.

Note here that, when the impurities are absent,
the present model reduces to two one-dimensional lattice models along the x- and 
y-axes, respectively.
In the present study the external force is applied 
mainly along the x-axis.
Some features of pinning and sliding may depend quantitatively
on its sliding direction to some extent and on the lattice symmetry.
However, the essential features in the presence of both periodic and random 
potentials will be unchanged, even though the direction of external force and the 
lattice symmetry are changed. 
That is, in the presence of both potentials, 
the frustration that causes the reduction of $F_c$ will occur generally, and 
the crossover of $F_c^y$ is expected to appear for any anisotropic periodic potential 
if the underlying periodic potential causes certain finite static pinning force 
originally in the transverse direction. 

In summary, we have investigated the static and dynamical behaviors of 
two-dimensional elastic lattices under periodic and impurity potentials
using a two-dimensional Frenkel-Kontorova model with impurities.
It is shown that the impurity can assist the depinning.
The crossover from moving Bragg glass states to moving transverse glass states is clearly seen 
in the structure factor.
The change of the lattice order affects the kinetic frictional force and 
the transverse depinning threshold force.
The latter shows a peculiar velocity dependence in the presence of 
both periodic and random potentials.
Such a transverse pinning effect may be observed also in experiments 
for a vortex lattice of type-II superconductor and 
vortices in a disordered Josephson junction array under a magnetic field.

One of the authors (H.M.) thanks V. M. Vinokur and M. Yumoto for valuable discussions.
The computation in this work was done using the facilities of the 
Supercomputer Center,
Institute for Solid State Physics, University of Tokyo and the facilities 
of the data processing center, Kyoto University.
This work was financially supported by
Grants-in-Aid for Scientific Research from the Ministry of Education,
Science, Sports and Culture.


%
%

\begin{center}
FIGURE CAPTIONS
\end{center}

\noindent
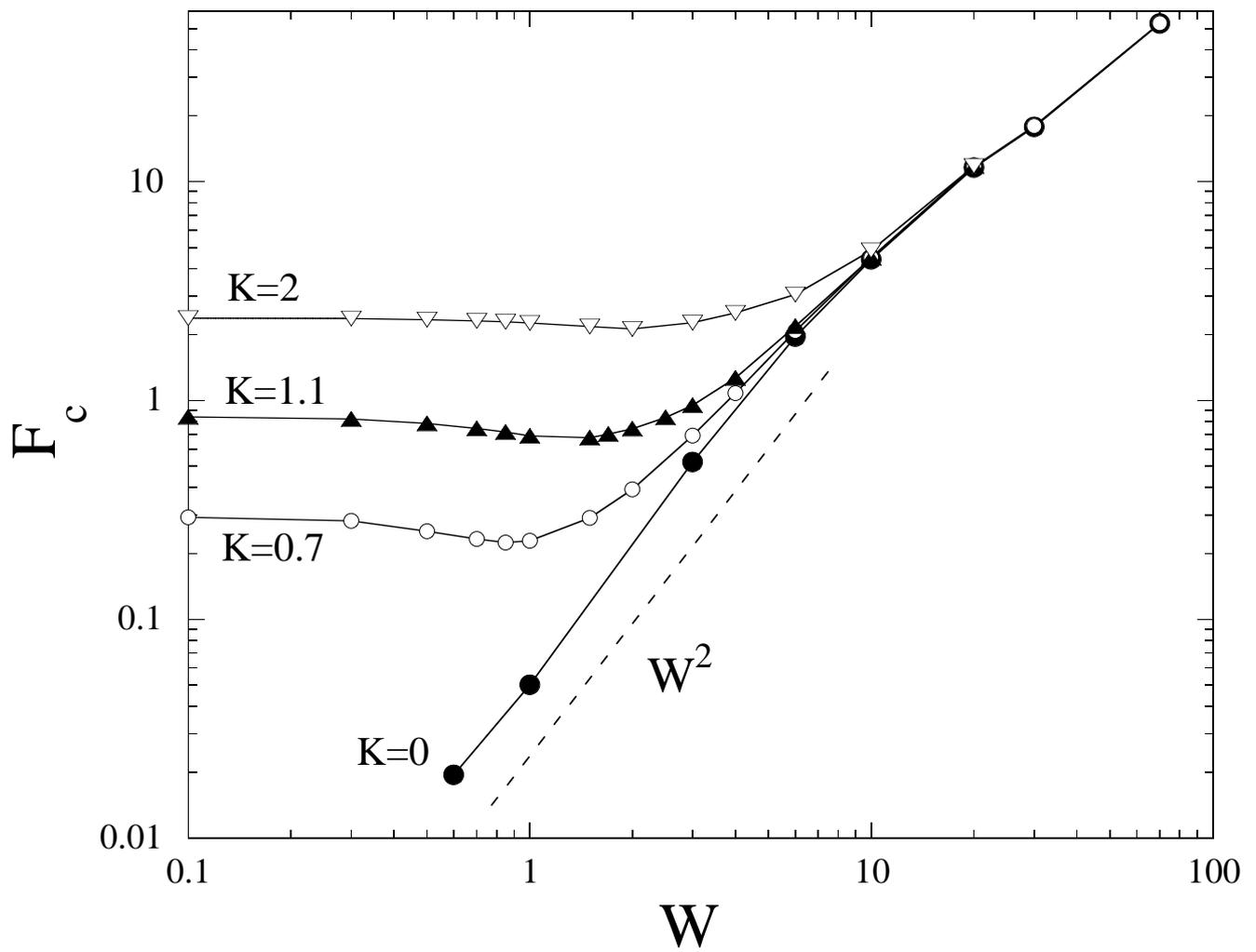
Fig. 1.
Longitudinal depinning threshold force $F_c$ versus the strength of 
impurity potential $W$,
for several strengths of periodic potential $K$'s.
\vspace{5mm}

\noindent
Fig. 2.
Structure factors in sliding states.
(a)$v_x= 0.96$, (b)$v_x= 31.6$, where $W=10$ and $K=0$.
\vspace{5mm}

\noindent
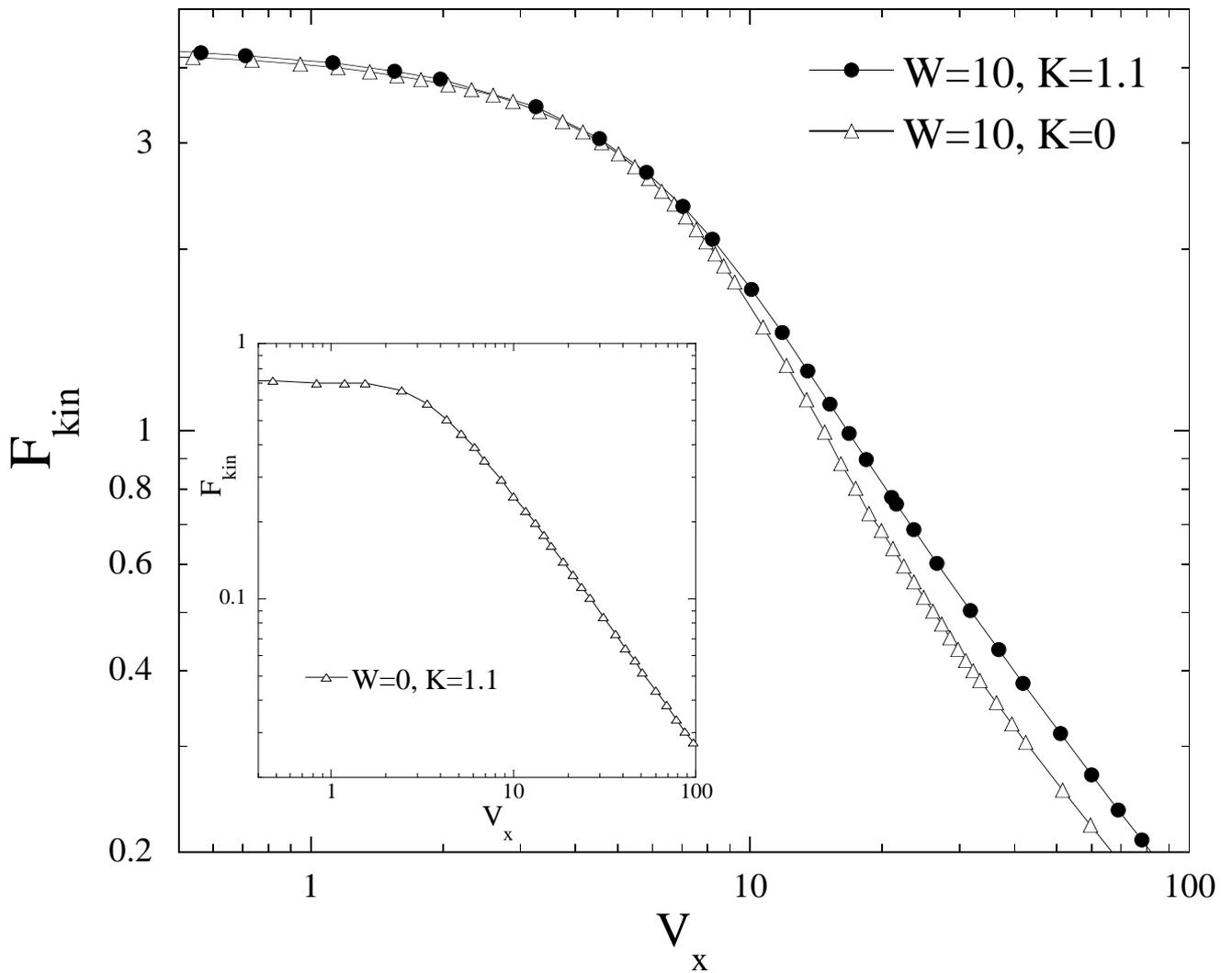
Fig. 3.
Kinetic frictional force $F_{kin}$ versus velocity $v_x$,
for several strengths of periodic potential $K$'s.
\vspace{5mm}

\noindent
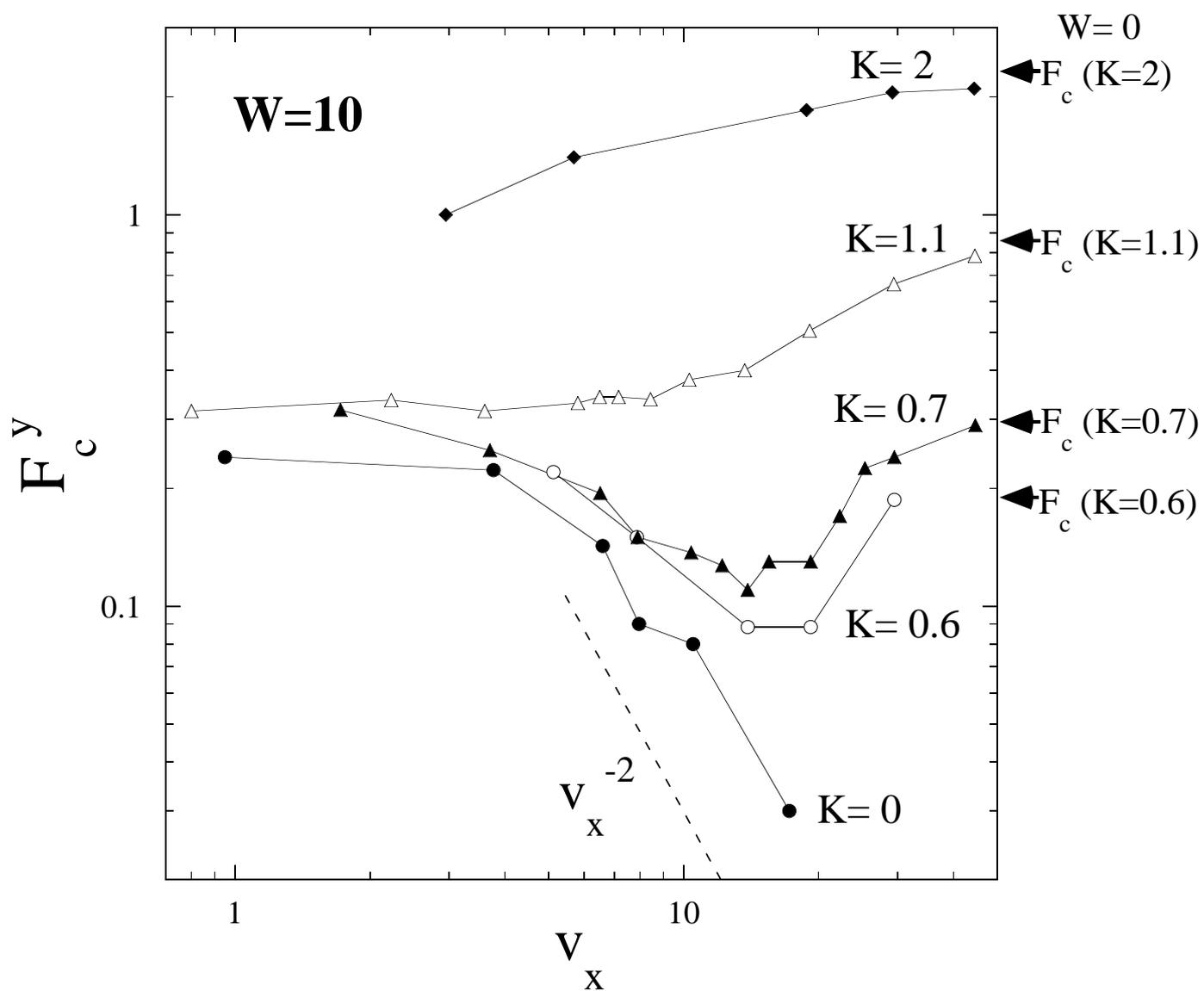
Fig. 4.
Transverse depinning threshold force $F_c^y$ versus longitudinal velocity 
$v_x$,
for several strengths of periodic potential $K$'s.
\vspace{5mm}

\end{document}